\documentclass[manuscript]{aa}

\usepackage{natbib}

\newcommand{\halfa}{H$_{\alpha}$}
\newcommand{\vzero}{V~0332+53}

\usepackage{graphicx}
\usepackage{amsmath}
\usepackage{amssymb}

\title{Activity from the Be/X-ray binary system V0332+53 during its intermediate-luminosity outburst in 2\,008}
\author{ M. D. Caballero-Garc\'{i}a
     \inst{1}
    \and
    A. Camero-Arranz
    \inst{2}
    \and
    M. \"{O}zbey Arabac{\i}
    \inst{3}
    \and
    C. Zurita                 
    \inst{4,5}
    \and
    J. Suso                   
    \inst{6}
    \and
    J. Guti\'{e}rrez-Soto     
    \inst{7,8}
    \and
    E. Beklen
    \inst{9}
    \and
    F. Kiaeerad               
    \inst{10,11}
    \and
    R. Garrido                
    \inst{8}
     \and
     R. Hudec
     \inst{12,1}
     }
     \institute{
$^1$ Czech Technical University in Prague, Faculty of Electrical Engineering, Prague, Czech Republic.\\
$^2$ Institut de Ci\`{e}ncies de l'Espai, (IEEC-CSIC), Campus UAB, Fac. de Ci\`{e}ncies, Torre C5 pa., 08193, Barcelona, Spain.\\
$^3$ Department of Physics, Middle East Technical University, Ankara, 06531, Turkey.\\
$^4$ Instituto de Astrof\'{\i}sica de Canarias, E-38200, La Laguna, Tenerife, Spain.  \\
$^5$ Universidad de La Laguna, Dept. Astrof\'{i}sica, E-38206, La laguna, Tenerife, Spain. \\
$^6$ Observatorio Astron\'{o}mico de la Univ. de Valencia, C/Catedr\'{a}tico Jose Beltran, 2, 46980 Paterna (Valencia), Spain. \\
$^7$ Universitad de Valencia, Dept. Did\'{a}ctica de las Matem\'{a}tica, Avda. Tarongers, 4, 46022, Valencia, Spain. \\
$^8$ Instituto de Astrof\'{i}sica de Andaluc\'{i}a (CSIC), Glorieta de la Astronom\'{i}a s/n, 18008, Granada, Spain.\\
$^9$ Physics Department, S\"{u}leyman Demirel University, 32260 Isparta, Turkey.  \\
$^{10}$ Nordic Optical Telescope, Apartado 474, 38700 Santa Cruz de La Palma, Spain.\\
$^{11}$ Department of Astronomy, Oscar Klein Center, Stockholm University, AlbaNova, Stockholm SE-10691, Sweden.\\
$^{12}$ Astronomical Institute, Academy of Sciences of the Czech Republic, 251~65~Ond\v{r}ejov, Czech Republic.\\
}  

\authorrunning{Caballero-Garc\'{i}a et al.}
\titlerunning{Activity from V~0332+53}
\offprints{M.~D. Caballero-Garcia\\  \email{cabalma1@fel.cvut.cz}}
\date{Received ; accepted}

\begin{document}

\abstract{}
{We present a study of the Be/X-ray binary system V 0332+53 with the main goal
of characterizing its behavior mainly during the intermediate-luminosity X-ray
event on 2008. In addition, we aim to contribute to the understanding of the global
behavior of the donor companion by including optical data from our dedicated
campaign starting on 2006. }
{{\vzero} was observed by \textit{RXTE} and \textit{Swift} during the decay of the
intermediate-luminosity X-ray outburst of 2008, as well as with \textit{Suzaku} before the rising 
of the third normal outburst of the 2010 series. In addition, we present recent data from the Spanish ground-based astronomical
observatories of El Teide (Tenerife), Roque de los Muchachos (La Palma), and Sierra
Nevada (Granada), and since 2006 from the Turkish T\"{U}B\.{I}TAK National Observatory
(Antalya). We have performed temporal analyses to investigate the transient behavior
of this system during several outbursts.  }
{Our optical study revealed that continuous mass ejection episodes from the Be star
have been taking place since 2006 and another one is currently
ongoing. The broad-band 1--60 keV X-ray spectrum of the neutron star during the
decay of the 2008 outburst was well fitted with standard phenomenological models, 
enhanced by an absorption feature of unknown origin at about 10\,keV and a narrow
iron K-alpha fluorescence line at 6.4\,keV. For the first time in {\vzero} we 
tentatively see an increase of the cyclotron line energy with increasing flux
(although further and more sensitive observations are
needed to confirm this). Regarding the fast aperiodic variability, we detect a 
Quasi-Periodic Oscillation (QPO) at $227{\pm}9$\,mHz only during the lowest 
luminosities. The latter fact might indicate that the inner regions surrounding the 
magnetosphere are more visible during the lowest flux states.  }
{}
{}

\maketitle

\section{Introduction}

Accreting X-ray pulsars are binary systems composed of a donor star and an accreting neutron star (NS). In High Mass X-Ray 
Binary (HMXB) systems the optical companion could be either a massive early-type supergiant (supergiant systems) or an 
O,B main sequence or giant star (BeX binaries; BeXB). Among the most remarkable signatures found in BeXBs are the detection of 
IR excess and emission-line features in their optical spectra produced in a disc-like outflow around the Be star. Historically, their 
outbursts have been divided into two classes. Type I (or normal) outbursts normally peak at or close to periastron passage of the 
NS (L$_X\leq$ 10$^{37}$\,erg s$^{-1}$). Type II (or giant) outbursts reach luminosities of the order of the Eddington 
luminosity (L$_X\sim$10$^{38}$\,erg s$^{-1}$; \citealt{frank02}), with no preferred orbital phase. We will refer as ``intermediate''
to any X-ray outburst that does not comply with the aforementioned properties, thus has (L$_X\sim$10$^{37}$-$10^{38}$\,erg s$^{-1}$).

\subsection{\vzero}

The recurrent hard X-ray transient {\vzero} was detected with the Vela 5B observatory in 1973 during its giant outburst, reaching a 
peak intensity of ${\approx}1.6$\,Crab in the 3--12\,keV energy band \citep{terrell84}. The system had passed a ten-year X-ray 
quiescent phase when 4.4\,s pulsations were detected with Tenma and EXOSAT satellites \citep{stella85}. These X-ray activities, a series 
of Type I outbursts, lasted about three months separated by the orbital period (34.25\,d) of the system. During these outbursts, rapid 
random fluctuations in the X-ray emission in addition to the pulse-profile variations were also reported \citep{unger92}. The system 
underwent another outburst, classified as Type II, in 1989 which led to the discovery of a cyclotron line scattering feature (CRSF) at 28.5\,keV 
and two QPOs centred at 0.051\,Hz and 0.22\,Hz \citep{makishima90,takeshima94,qu05}.

The optical companion of the system, BQ~Cam \citep{argyle83} is an O8-9 type main sequence star at a distance of ${\approx}$7\,kpc \citep{negueruela99}. It 
has been widely observed both in optical and IR wavelengths since its identification. The optical spectrum is characterised by the highly variable 
emissions of ${\rm H}_{\alpha}$, ${\rm H}_{\beta}$ and ${\rm H}_{\gamma}$ in addition to the He\,I lines. The photometric data shows 
an IR excess \citep{bernacca84,coe87,honeycutt85,unger92}. The brightening in optical/IR light-curves is 
usually accompanied with the X-ray outburst phases as in the case of giant 2004 outburst of the system. \citet{goranskij04} predicted this 
outburst based on the optical brightening of {\vzero} in optical/IR band. During the outburst, three additional CRSFs at 27, 51 and 74\,keV were 
detected in {\it Rossi X-ray Timing Explorer} ({\it RXTE}) observations \citep{coburn05} and confirmed by the subsequent {\it INTEGRAL} data 
\citep{pottschmidt05}. \citet{tsygankov06} showed that the energy of these features are linearly anti-correlated with its luminosity. 
The following X-ray activities of the system were in 2008, 2009 and 2010 with relatively weaker peak fluxes 
\citep{krimm08,krimm09,nakajima10}. The system was in X-ray quiescence until 18 June 2015, when the Be companion probably reached its 
maximum optical brightness after renewing activity in 2012 \citep{cameroarranz15}.  

Here, we present a multi-wavelength study of {\vzero} mostly during the recent events initiated in 2008. For this purpose we used
archived {\it Swift}-XRT and {\it RXTE} pointed observations carried out in 2008, as well as one {\it Suzaku} observation from 2010
and survey data from different space-borne telescopes and covering intermediate and low-luminosity events. In addition, we used optical/IR data from
our dedicated campaign involving several ground-based astronomical observatories. This provides us a unique opportunity to undertake studies
of X-ray outbursts of intermediate and low X-ray luminosities. The latter is a regime very scarcely explored in {\vzero}, one of the BeX with strongest
magnetic field already known.

\section{Observations and  Data Reduction}\label{observations}

\subsection{Optical/IR photometric observations}\label{photobs}

\begin{figure}
\centering
 \includegraphics[bb=-25 50 554 770,width=7.5cm,angle=270,clip]{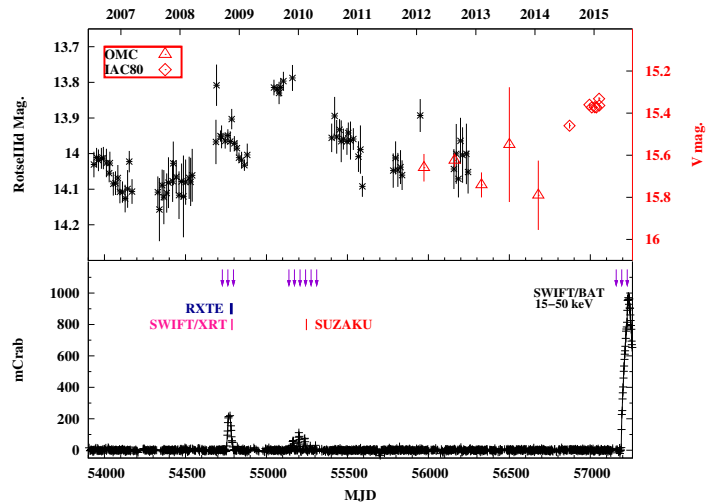}
 \caption{Comparison of the {\it Swift}/BAT light curve (15-50\,keV) 
with the optical magnitudes (ROTSE, OMC, IAC80) of {\vzero} since 2006. The BAT, ROTSE and OMC data points have 
been rebinned for clarity. The violet arrows in the X-ray panel represent the times of the periastron 
passages of the NS. }
 \label{fig1}
\end{figure}

The optical counterpart has been observed in the optical band with the 80-cm IAC80
telescope at the Observatorio del Teide (OT; Tenerife, Spain) starting on August
2014 until February 2015 (see on-line material Tab.~\ref{otab1} and Fig.~\ref{fig1}). We 
obtained the optical photometric CCD images using B and V filters with integration time of
60\,s. In addition, on 20 December 2014 a single infrared J, H and K$_{\rm s}$
observation was performed by the CAIN camera from the 1.5\,m TCS telescope at the OT,
with integration time of 150, 90 and 90\,s, respectively. The reduction of the data was done by using the
pipelines of both telescopes based on the standard aperture 
photometry \citep[for more details on reduction]{camero14}.

The main part of the long-term optical CCD observations of {\vzero} include data
from the 0.45\,m ROTSEIIId\footnote{The Robotic Optical Transient Search Experiment, ROTSE, is a
collaboration of Lawrence Livermore National Lab, Los Alamos National Lab, and the University of
Michigan (http://www.ROTSE.net)} telescope from the T\"{U}B\.{I}TAK National Observatory
(TUG; Antalya, Turkey) between February 2006 until December 2012. This telescope
operates without filters and is equipped with a 2048 $\times$ 2048 pixel CCD. Dark
and flat-field corrections of all images were done automatically by a pipeline. 
Instrumental magnitudes of all the corrected images were obtained using the
SExtractor Package \citep{bertin96}. Calibrated ROTSEIIId magnitudes were acquired
by comparing all the stars in each frames with USNO-A2.0 catalog $R$--band
magnitudes.

In addition, we have used data from the Optical Monitoring Camera (OMC; \citealt{mashesse03}) 
on board the {\it INTEGRAL} satellite \citep{winkler03}. They were obtained from the
Data Archive Unit at CAB Astrobiology centre\footnote{http://sdc.cab.inta-csic.es/omc/index.jsp} (INTA-CSIC), and
pre-processed by the INTEGRAL Science Data Centre (ISDC)\footnote{http://www.isdc.unige.ch}.

\subsection{Optical spectroscopic observations}  \label{specobs}

Optical spectroscopic observations of the donor star were performed since 26
September 2006 until 22 November 2014 (see on-line material Tab.\ref{otab2}) with 
three different telescopes: the Russian-Turkish 1.5\,m telescope (RTT150) at the 
TUG, the 2.56\,m Nordic Optical Telescope (NOT) located at the Observatorio del
Roque de los Muchachos (La Palma, Spain), and the 1.5\,m Telescope at the
Observatorio de Sierra Nevada (OSN-CSIC) in Granada (Spain).

The spectroscopic data from RTT150 were obtained with the T\"{U}B\.{I}TAK Faint
Object Spectrometer and Camera (TFOSC). The reduction of the spectra was done using the Long-Slit
package of  MIDAS\footnote{http://www.eso.org/projects/esomidas}. The
low-resolution OSN spectra were acquired using Albireo spectrograph whereas the NOT
spectra were obtained with the Andaluc\'ia Faint Object Spectrograph and Camera
(ALFOSC). The reduction of this data set was performed  using standard procedures
within IRAF\footnote{IRAF is distributed by the National Optical Astronomy
Observatory, optical images which is operated by the Association of Universities for
Research in Astronomy (AURA) under cooperative agreement with the National Science
Foundation.} (\citealt{camero14}, for further details on the instrumentation and
data reduction).

All spectroscopic data were normalized with a spline fit to continuum and corrected
to the barycenter after the wavelength calibration. The full width at half maximum (FWHM) 
and equivalent width (EW) measurements of H$_\alpha$ lines were acquired by fitting Gaussian 
functions to the emission profiles using the ALICE subroutine of MIDAS.

\begin{table}
 \centering
 \begin{minipage}{120mm}
  \caption{Log of the X-ray observations\,$^{1}$.}
  \label{tab2}
  \begin{tabular}{@{}lcccll@{}}
  \hline \noalign{\smallskip}
   Obs.  &    Date\,$^2$                 &   MJD                         &          Telescope     & Exp.                    &    Pulse                   \\
         &                               &                               &                        &          time           &    period                   \\
         &                               &                               &                        &               (s)       &               (s)           \\
 \hline   \noalign{\smallskip}
  1   &     04/11/08                &   54774.636                 &   {\it RXTE}\,$^{3}$   &  1\,600                 &   4.3742(1)                  \\
  2   &     07/11/08                &   54777.999                 &   {\it RXTE}           &  752                    &   --                         \\
  3   &     09/11/08                &   54779.731                 &   {\it RXTE}           &  496                    &   --                         \\
  4   &     10/11/08                &   54780.594                 &   {\it RXTE}           &  864                    &   --                         \\
  5   &     12/11/08                &   54782.394                 &   {\it RXTE}           &  1\,456                 &   4.3740(1)                  \\
  6   &     14/11/08                &   54784.511                 &   {\it RXTE}           &  1\,552                 &   4.3746(6)                  \\
  7   &     16/11/08                &   54786.543                 &   {\it RXTE}           &  1\,488                 &   4.37450(2)                 \\
  8   &     19/11/08                &   54789.826                 &   {\it RXTE}           &  3\,840                 &   4.37540(2)                 \\
  9   &     20/11/08                &   54790.775                 &   {\it RXTE}           &  2\,848                 &   4.3763(1)                  \\
 10   &     21/11/08                &   54791.743                 &   {\it RXTE}           &  1\,152                 &   4.3760(1)                  \\
 11   &     22/11/08                &   54792.270                 &   {\it RXTE}           &  3\,716                 &   4.37610(4)                 \\
 12   &     23/11/08                &   54793.273                 &   {\it RXTE}           &  1\,597                 &   4.37620(4)                 \\
 13   &     23/11/08                &   54793.387                 &   {\it RXTE}           &  3\,235                 &   4.37620(5)                 \\
 \hline   \noalign{\smallskip}
 (5)  &     12/11/08                &   54782.384                 &   {\it Swift}\,$^{4}$  &  1\,706             &   --                         \\
 (6)  &     14/11/08                &   54784.520                 &   {\it Swift}          &  2\,108                 &   --                         \\
 (7)  &     16/11/08                &   54786.533                 &   {\it Swift}          &  2\,307                 &   --                         \\
 (8)  &     19/11/08                &   54789.338                 &   {\it Swift}          &  1\,613                 &   --                         \\
 (9)  &     20/11/08                &   54790.811                 &   {\it Swift}          &  1\,597                 &   --                         \\
 (10) &     21/11/08                &   54791.487                 &   {\it Swift}          &  1\,489                 &   --                         \\
 \hline   \noalign{\smallskip}
 14   &     16/02/10          &   55243.264                 &   {\it Suzaku}\,$^5$   &  15\,946                &   4.37(1)                   \\
\noalign{\smallskip}
\hline  \noalign{\smallskip}
\multicolumn{6}{l}{$^1$ Standard mode for {\it RXTE} and {\it Suzaku} and Windowing }\\
\multicolumn{6}{l}{mode for {\it Swift}-XRT. }\\
\multicolumn{6}{l}{$^2$ Observation date in the format (dd/mm/(20)yy). }\\
\multicolumn{6}{l}{$^3$ {\it RXTE}/PCA exposure time. }\\
\multicolumn{6}{l}{$^4$ {\it Swift}/XRT exposure time. }\\
\multicolumn{6}{l}{$^5$ {\it Suzaku}/XIS1 exposure time. }\\
\end{tabular}
\end{minipage}
\end{table}

\begin{figure*}
\centering
 \includegraphics[bb=0 0 680 340,width=18.7cm,angle=0,clip]{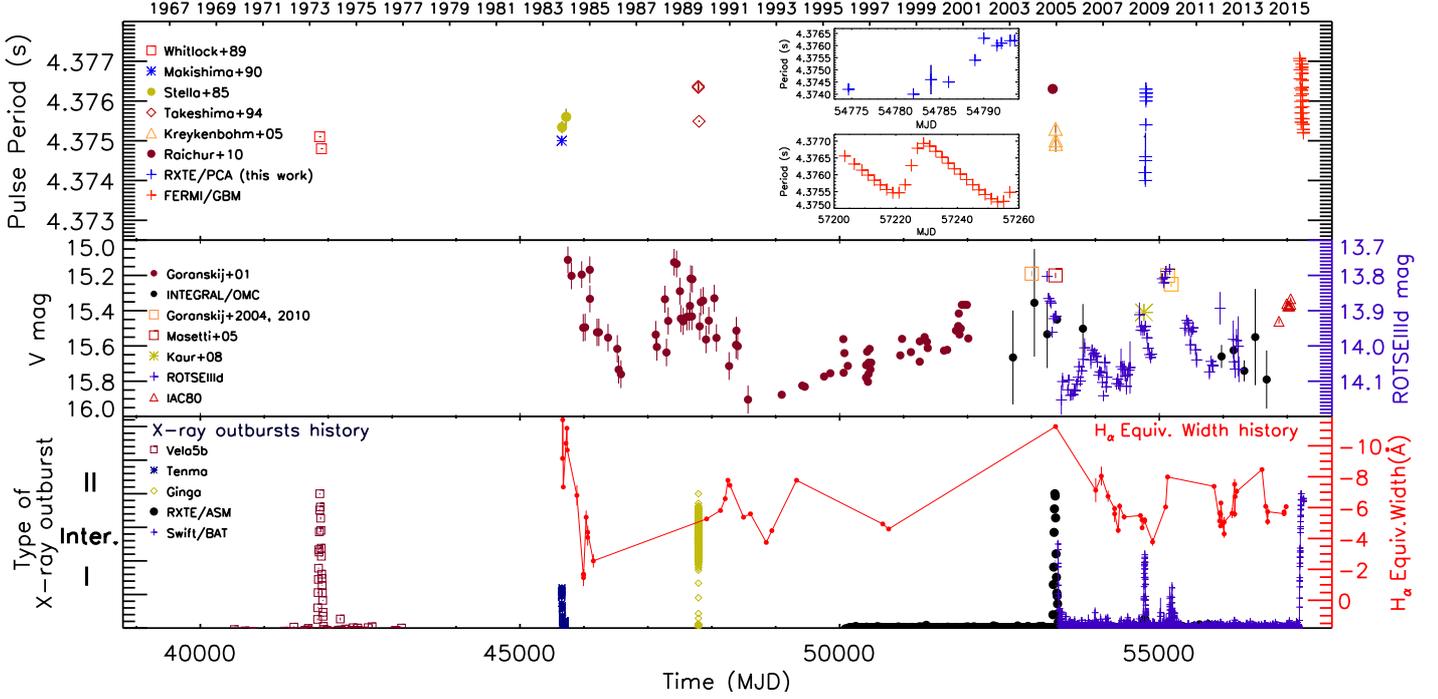}
 \caption{\textbf{Top}. Long-term pulse period history of {\vzero} since its discovery in 1973. \textbf{Middle}. Long-term optical lightcurve of {\vzero}. All but the 
IAC80 light curve have been rebinned for clarity. \textbf{Bottom}. Type of X-ray outburst underwent by {\vzero} (i.e. Type II,
Intermediate, and Type I). The intensities are the same for outbursts of the same type, not having been corrected for the different energy bands. We only intend to be
illustrative of the times and types of events. The long-term evolution of the {\halfa} EW
(red filled circles) is also overplotted. The measurements have been extracted from \citet{iye85}, \citet{kodaira85}, \citet{corbet86}, \citet{negueruela99}, \citet{masetti05}, \citet{kiziloglu08}, \citet{goranskij10} and the present work. 
 } 
 \label{fig7}
\end{figure*}

\subsection{X-ray Observations}\label{obs}

The XRT (on-board the {\it Swift} satellite; \citealt{gehrels04}) observed {\vzero}
in Windowing Timing (WT) mode (1.7\,ms time resolution, 1D Image) six times from 12 to
21 November 2008. Tab.~\ref{tab2} shows a summary of the log of these observations. To 
reduce the data and extract final products (spectra and light curves) we
followed the standard recipe as detailed in the {\it Swift}/XRT data reduction
guide\footnote{http://swift.gsfc.nasa.gov/analysis/xrt\_swguide\_v1\_2.pdf}. The
relevant response matrix to use is given by the HEASARC's calibration database
(CALDB\footnote{http://heasarc.gsfc.nasa.gov/FTP/caldb}). All the XRT lightcurves
were barycentred corrected. We also used for comparison data products supplied by
the UK \textit{Swift} Science Data Centre at the University of Leicester
\citep{evans09}. 

{\it RXTE} {\it Proportional Counter Array} (PCA; \citealt{jahoda96}) standard 1 and
2 data only from PCU-2 were selected for this work. To extract PCA products we
followed the standard procedure showed in its
cookbook\footnote{https://heasarc.gsfc.nasa.gov/docs/xte/recipes/\\pca\_spectra.html}. A 
systematic error of 0.5$\%$ was added in quadrature to the PCA standard 2
spectra, as recommended by the instrument team. Products from the High Energy X-ray
Timing Experiment (HEXTE; \citealt{gruber96}) were also extracted using the corresponding
cookbook recipe\footnote{https://heasarc.gsfc.nasa.gov/docs/xte/recipes/hexte.html}.
Only standard mode data from cluster B were used. All the {\it RXTE} lightcurves
were barycentred corrected.

In addition, we analysed one observation performed by {\it Suzaku} \citep{mitsuda07}
in 16 February 2010. We selected only data from the X-ray Imaging Spectrometer (XIS;
\citealt{koyama07}) CCD and the PIN diode detector (10-70\,keV) of the he Hard X-ray
Detector (HXD; \citealt{takahashi07}). For this observation only XIS0, XIS1 and XIS3
detectors were available. Following the {\it Suzaku ABC
guide}\footnote{http://heasarc.nasa.gov/docs/suzaku/analysis/abc}, we reprocessed
all the data and extracted the final products. The clocking mode option was normal
burst (1.0) with the 1/4 window selected (1\,s time binning), and a data editing
mode of 3${\times}$3. No pile-up correction was needed. The events were barycentred
corrected.

We used the HEAsoft software version 6.15.1 \citep{arnaud96} for the data analysis
of all the instruments. Since 2008, the Gamma-ray Burst Monitor (GBM) on board the
\textit{Fermi} satellite, has been monitoring {\vzero}. In this work we used timing
products provided by the GBM Pulsar
Team\footnote{http://gammaray.nsstc.nasa.gov/gbm/science/pulsars} \citep[see
e.g.][for a detailed description of the timing technique]{finger09,camero10}. We
also used quick-look X-ray results provided by the \textit{RXTE} All Sky Monitor
team\footnote{http://xte.mit.edu/ASM$\_$lc.html}, and
since 2008 we used the quick-look X-ray \textit{Swift}/BAT transient monitor results
provided by the \textit{Swift}/BAT
team\footnote{http://swift.gsfc.nasa.gov/results/transients} \citep{krimm13}. See
Fig.~\ref{fig7}.

\section{Results}

\subsection{Optical photometry}

The optical light-curves of BQ~Cam compared to the {\it Swift}/BAT data (15-50\,keV) from June 2006 to June 2015 are shown in Fig.~\ref{fig1}. In 2008, after being ${\approx}3.5$\,yrs
in quiescence in X-rays the system underwent a new active period. However the optical companion entered the brightening state nearly 1.5 yr before the NS. When
the enhancement in the optical brightness reached a value of ${\approx}0.25$\,mag. The X-ray outburst was triggered, around 17 October 2008 (MJD 54756) roughly six days before the
periastron passage of the NS (vertical red-dashed lines in Fig.~\ref{fig1}). This large flare lasted about 40\,d and reached a flux of ${\approx}$214\,mCrab within 3 weeks. It 
should be pointed out that the optical magnitude of {\vzero} shows a sudden decrease when the X-ray outburst declines. Fading of the optical companion continued until the end 
of January 2009 (MJD ${\approx}$54850) and soon after it turned back to its brightening state. The X-ray flux, on the other hand, stayed at its quiescent level, until the Be star 
reached a peak value of ${\approx}$13.8\,mag (ROTSE magnitude). A new X-ray phase of the system started
in November 2009 again a few days after the periastron passage (MJD ${\approx}$55140) of the NS. That activity included five small but periodic outbursts separated by the
orbital period (34.25\,d) of the system classified as Type I. 

The second outburst of these series was the most powerful one with a peak flux of ${\approx}$110\,mCrab. The
optical magnitude, on the other hand, started to decrease with declining of the second outburst (MJD ${\approx}$55201). The X-ray activities finished by the end of the
May 2010 (MJD ${\approx}$55312) while the optical magnitude was still fading. It ceased around August 2011 (MJD ${\approx}$55780) and lasted ${\approx}$4 years. 
Our recent data from the IAC80 telescope and the optical camera OMC on board {\it INTEGRAL}, shows an optical enhancement of the counterpart BQ~Cam, with the brightness varying 
from the quiescence level of ${\rm V}=15.93{\pm}0.55$ (06 Jan. 2012; MJD 55932.60) to ${\rm V}=15.332{\pm}0.008$ (1 Feb. 2015; MJD 57054.04; \citealt{cameroarranz15}). This corresponds to an increment 
of ${\approx}0.6$\,mag, similar to that observed during the brightening episodes occurred e.g. in 1983 (see \citealt{goranskii01}) and in 2004 \citep{goranskij04}, but higher 
than the ones in 2008 \citep{kaur08} and 2009 \citep{goranskij10}. This is confirmed by our only IR measurement from December 2014 (MJD 57011.869), that showed an unusually bright Be companion. The 
magnitudes obtained were J=11.228$\pm$0.003\,mag, H=10.651$\pm$0.003\,mag, and ${\rm K}_{\rm s}$=10.285$\pm$0.004\,mag, which are comparable to those from December 1983 
\citep{williams83}. We note that BQ~Cam did not reach its maximum level (${\rm V}=15.2$\,mag) in our observations from February 2015. The maximum has been 
probably reached recently, since the BAT monitor is currently detecting X-ray activity from {\vzero} with a daily average flux of the order of 310.4\,mCrab on 25 June 2015 (MJD 57198.0).

\subsection{${\rm H}_{\alpha}$ line}

\begin{figure*}
\centering
 \includegraphics[bb=28 28 570 810,width=13.0cm,angle=270,clip]{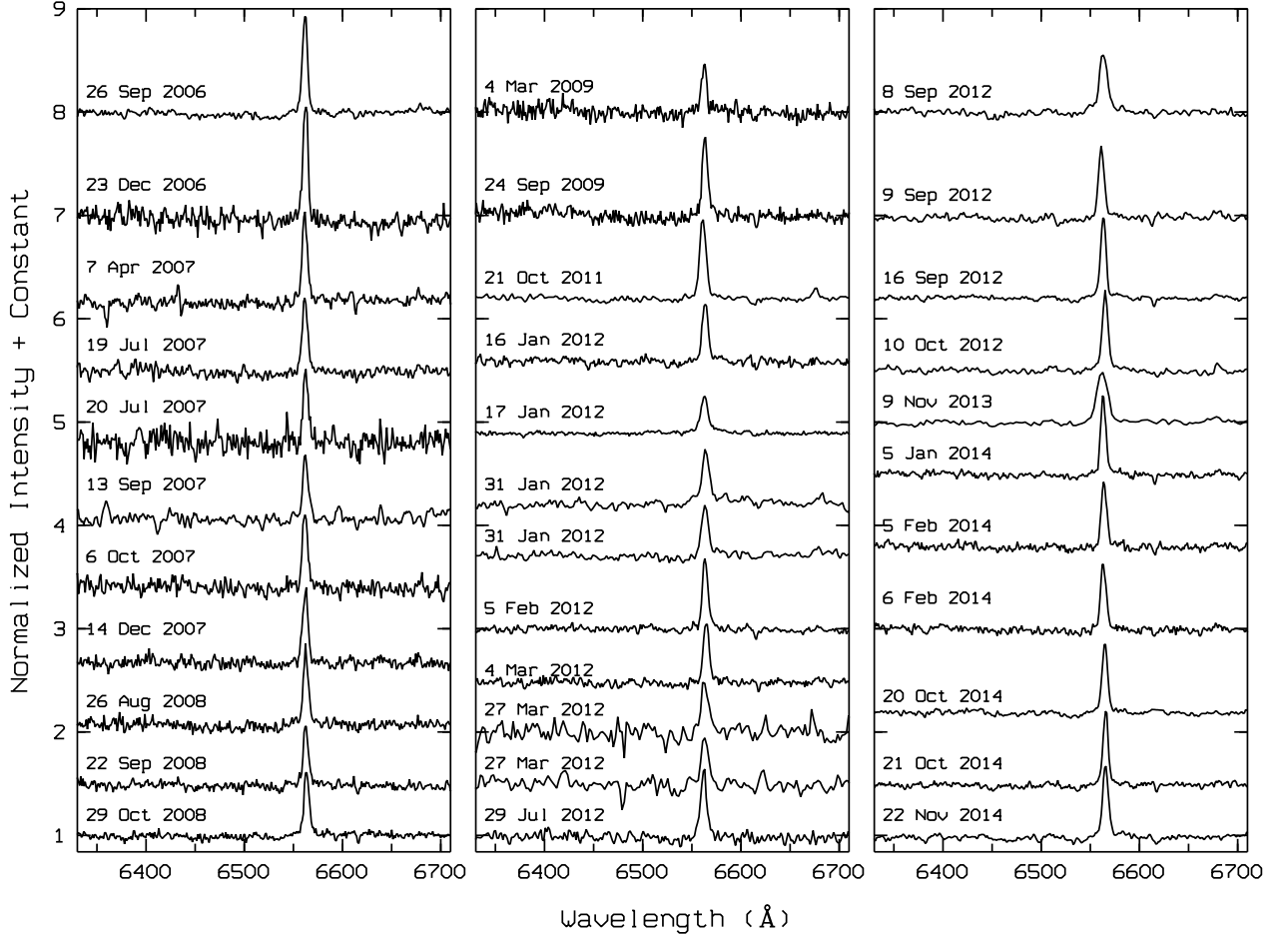}
 \caption{${\rm H}_{\alpha}$ profiles of {\vzero} selected between 2006 and 2014. The single-peaked emission of the line does not change on a long-term basis.  }
 \label{fig2}
\end{figure*}

The long-term ${\rm H}_{\alpha}$ line monitoring of BQ~Cam from September 2006 to November 2014 has been studied (see Fig.~\ref{fig2}). In contrast to the previous works stating the
${\rm H}_{\alpha}$ profile variations \citep{negueruela98,negueruela99}, we did not see such variability patterns during the observation period. Instead, all the line
profiles are nearly-symmetric in a single-peaked form despite the lack of the observations for the period 2010-2011.

\begin{figure}
\centering
 \includegraphics[bb=50 50 554 770,width=7.7cm,angle=270,clip]{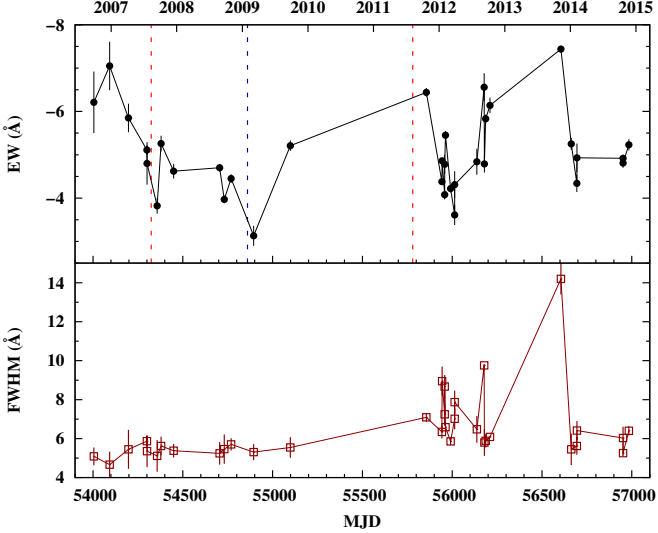} \caption{Evolution since 2006 of the EW and FWHM measurements of the ${\rm H}_{\alpha}$ emission line. The red-dashed vertical line denotes the beginning and the end of the optical
outburst seen in ROTSEIIId light curve (see also Fig.~\ref{fig1}) whereas the black-dashed line shows the time of the turning point from decreasing to increasing trend just
after (${\approx}10$\,d) the strong X-ray outburst around MJD${\approx}$54850. }
 \label{fig3}
\end{figure}

In spite of having the same single-peaked emission profile on a long-term basis, the ${\rm H}_{\alpha}$ line strength does not show a constant trend during the
spectroscopic observations (see on-line material Tab.~\ref{otab2}). In Fig.~\ref{fig3} the evolution of EW and FWHM measurements of the emission line are shown. The
first two measurements of the line coincide with the declining phase of the optical outburst seen in the time interval MJD 53576-54180. The
EW values decreased as the star faded away which can be attributed to the weakening of the decretion disc. Besides, this decreasing pattern
continues after the optical outburst triggered (MJD 54324). Although our data coverage is not uniform, we see a reverse relation between the optical magnitude and
the width of the emission line until 4 March 2009 (MJD 54894) when the lowest value of the EW of ${\rm H}_{\alpha}$ is reached (see Fig.~\ref{fig7}). After about 
seven months, the EW reached a value
of -5.21\,{\rm \AA}. Although this sudden increase in the line width well fits with the rising of the optical magnitude, we cannot make any further explanation regarding
the behavior of the emission line due to the lack of data for 2010-2011. In 2012-2014, we see that the strength of the ${\rm H}_{\alpha}$ line kept its varying
pattern without a noticeable change in the optical magnitude indicating the upcoming mass ejection event.

In contrast to the variations in EW values, the FWHM values stayed approximately constant until the end of 2012. At that point, the FWHM values started to
follow nearly the same pattern of EW. Indeed, showing a similar behavior is not unusual for EW and FWHM, since the width of the emission profile is directly 
related to the rotational velocity of the Be star. \citet{hanuschik89} gives the relation between the ${\rm H}_{\alpha}$ widths
and the projected rotational velocity, $v{\sin(i)}$. The projected rotational velocity of {\vzero} is found to be $v{\sin(i)}=166.68\,{\rm km}\,{\rm s}^{-1}$, using
the average values of -5.04\,{\rm \AA} (i.e. $230\,{\rm km}\,{\rm s}^{-1}$) and 7.3\,{\rm \AA} ($333.56\,{\rm km}\,{\rm s}^{-1}$) for the EW and FWHM, respectively
(see on-line material Tab.~\ref{otab2}). For the true 
rotational velocity of {\vzero}, we have assumed the inclination angle of the system of ${\rm i}{\le}18.9^{\circ}$ as suggested by \citet{zhang05}. Therefore, the 
lower limit to the true rotational velocity would be $514\,{\rm km}\,{\rm s}^{-1}$. Using the typical mass and radius values for a late O-type 
star (${\rm M}{\gtrsim}20\,{\rm M}_{\odot}$ and ${\rm R}{\gtrsim}15\,{\rm R}_{\odot}$), the
break-up velocity is determined as $600\,{\rm km}\,{\rm s}^{-1}$ that is close to the rotational velocity.

\subsection{X-ray activity}

\subsubsection{Periodic timing analysis}

To search for periodic pulsations from the compact object we have used the tool {\tt efsearch} from the 
Xronos\footnote{http://heasarc.gsfc.nasa.gov/docs/xanadu/xronos/\\xronos.html} package. This tool searches for periodicities in a time series by folding the data 
over a range of periods and by searching for a maximum chi-square as a function of period. Standard\,1 PCA, XRT and {\it Suzaku}-XIS barycentred lightcurves for {\vzero} 
were selected for this task (see Tab.~\ref{tab2}). The resulted pulse periods can be seen in Fig.~\ref{fig7} and Tab. 1. The error on the measured spin periods is 
calculated using the chi-square versus spin period plot provided by {\tt efsearch}. The points near the peak of this plot are fitted with a gaussian. The error
on the gaussian centre estimate is taken as the error on the spin period. From Tab.~\ref{tab2} we can see that during the 2008 outburst decay the 
neutron star seems to spin slower, as expected. Additionally, we note that in Fig.~\ref{fig7} we do not include the pulse period obtained with {\it Suzaku}-XIS 
(${\rm P}_{pulse}=$4.37(1)\,s) because of its large uncertainty. Furthermore, we do not include any pulse period from the XRT data because they were not
significantly detected.

\subsubsection{Aperiodic timing analysis}

We performed an analysis of the fast time variability of {\vzero} in the 2-60\,keV energy range of Obs.~1,~2,~6,~8 and 9. The net PCA count
rate is ${\approx}700,600,300,300,160\,{\rm cts}\,{\rm s}^{-1}$ for Obs.~1-9, respectively. The rest of the observations either are too short and/or the source
is too faint to perform timing analysis. The time resolution of the PCA in the mode used is 0.125\,s.

We used the GHATS package, developed 
under the IDL environment at INAF-OAB\footnote{http://www.brera.inaf.it/utenti/belloni/\\GHATS\_Package/Home.html}, to produce the Power Density Spectra (PDS) from 
512 points in each light curve. The PDS were then averaged together for each observation. The PCA light curve was binned at its
time resolution (0.125\,s). This yields a Nyquist frequency of $=4$\,Hz. The PDS were normalized according to \citet{leahy83}. All of the PDS show low-frequency 
band-limited noise (that may turn into peaked noise) and QPO noise.

PDS fitting was carried out with the standard XSPEC fitting package by using a unit response. Fitting
the (2-60\,keV) PDS with a model constituted by a zero centred Lorentzian for the flat-topped noise (hereafter called ${\rm L}_{1}$) plus a constant for the Poissonian noise 
(i.e. well reproduced by a {\tt powerlaw} model component with its photon index frozen to zero in the fits) results in
mostly acceptable chi-square values only for the PDS of Obs.~1, (see Tab.~\ref{tab6}). In the case of Obs.~8,~9 the fit statistics with this
model is worse and leaves some visible residuals, i.e. ${\chi}^{2}/{\nu}=290/252,233/252$, respectively. The fit to these observations improved
by adding a further Lorentzian component (${\rm L}_{2}$) centred at ${\nu}=30-80\,$mHz. This component changed the (F-test) fits statistics by
${\Delta}{\chi}^{2}{\approx}20-30$ for ${\nu}=249$ d.o.f., thus a ${\le}3{\sigma}$ improvement. For Obs.~2,~6 we took into account the latter feature only, since
it is broad and therefore constitute a main part of the measure of the rms, i.e. by $(10-20)\%$. This resulted in a fit statistics of ${\chi}^{2}/{\nu}=253/252,255/252$
for Obs.~2,~6, respectively.

The fit of the (2-60\,keV) PDS with a model constituted by a zero (${\rm L}_{1}$) and a ${\nu}=30-80\,$mHz (${\rm L}_{2}$) centred Lorentzians (${\rm L}_{2}$ only for Obs.~2 and 6) 
for the band-limited noise plus a constant for the Poissonian noise results
in a poor description of the data for Obs.~6,~8,~9. The fit statistics with this model is of ${\chi}^{2}/{\nu}=255/252,263/249,209/249$ for Obs.~6-9, respectively. In
these observations there are positive residuals centred in the range of $(220-240)$\,mHz that we fitted by adding a further Lorentzian component (${\rm L}_{\rm QPO}$) centred
at those frequencies. This component changed the fits statistics by ${\Delta}{\chi}^{2}{\approx}12,19,13$ for ${\nu}=249,246,246$ d.o.f., thus a $(2.9,3.9,3.0){\sigma}$ 
improvement for Obs.~6-9, respectively. We took into account this feature, since it
is broad and therefore affects substantially the measure of the rms, i.e. by ${\approx}(2-4)\%$ for all the observations. These positive features have a peaked form, thus we will call
them ${\rm QPOs}$, consistently with previous findings \citep{qu05,reig08,reig13}. The detection of this peak is only significant for Obs.~8, which can be explained since this is the
observation with the longest exposure time of the sample.

\begin{figure*}
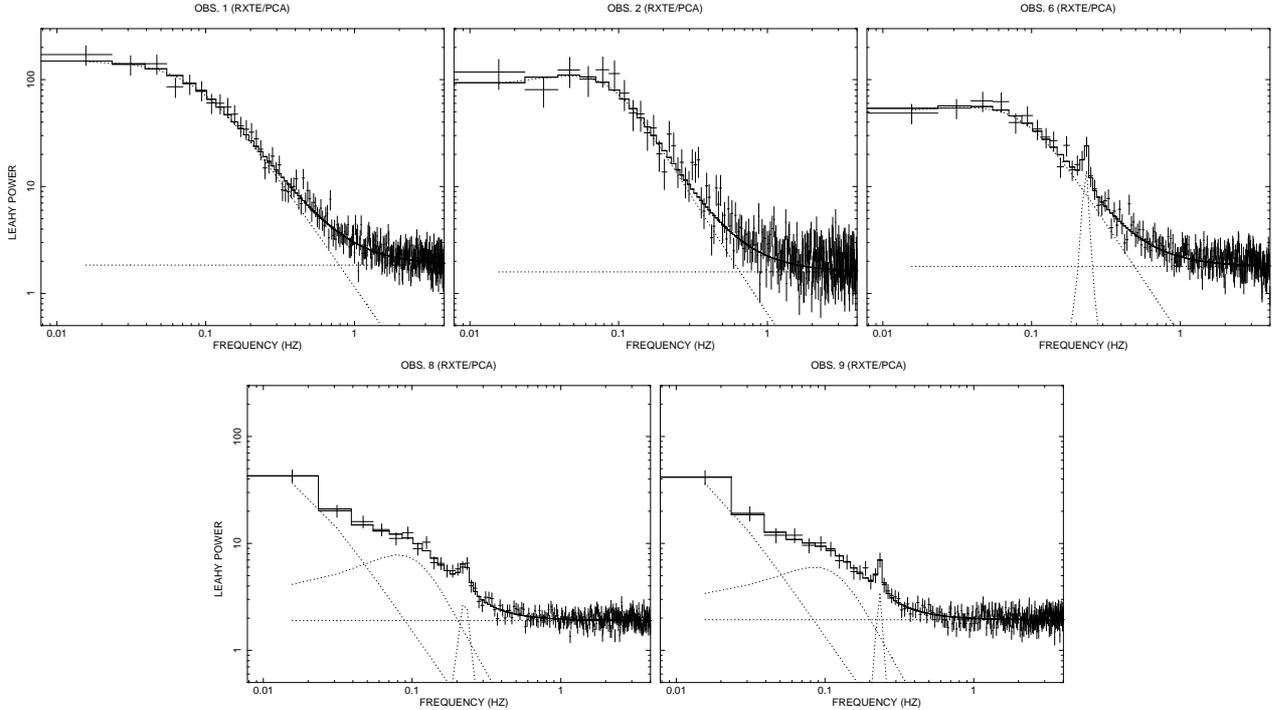

\centering
 \includegraphics[bb=34 5 570 700,width=4.7cm,angle=270,clip]{pds1.ps}
 \includegraphics[bb=34 93 570 700,width=4.7cm,angle=270,clip]{pds2.ps}
 \includegraphics[bb=34 93 570 700,width=4.7cm,angle=270,clip]{pds3.ps}
 \includegraphics[bb=34 5 570 700,width=4.7cm,angle=270,clip]{pds4.ps}
 \includegraphics[bb=34 93 570 700,width=4.7cm,angle=270,clip]{pds5.ps}
 \caption{Power density spectra of the {\it RXTE} PCA data in the energy and frequency range (2-60\,keV) and 0.008-4\,Hz, respectively, during observations 1, 2, 6, 8 and 9 (top-left to bottom-right) with the best fit model (solid line) and the model components (dotted line). The results of the fitting are shown in Tab.~\ref{tab6} and discussed in the text.  }
 \label{fig6}
\end{figure*}

We calculated the fractional rms from the best fit model (integrated in the $(8{\times}10^{-3}-4)$\,Hz band). This was found to be $(15-20)\%$ in
the 2--60\,keV energy range. The results obtained from the global fits (i.e. using the {\tt lorentz+lorentz+lorentz+powerlaw} model in XSPEC) are reported in Tab.~\ref{tab6}. We 
plot in Fig.~\ref{fig6} the broad-band PDS
with the best-fit model. We notice that our results are broadly in agreement with those previously obtained \citep{reig13}. 

\begin{table*}
 \centering
 \begin{minipage}{140mm}
  \caption{Results from the {\it RXTE} timing analysis$^{1}$.}
  \label{tab6}
  \begin{tabular}{@{}lccccc@{}}
  \hline  \noalign{\smallskip}
                                &   Obs.~1           &          Obs.~2        & Obs.~6                  &    Obs.~8             &    Obs.~9      \\
 \hline  \noalign{\smallskip}
   ${\nu}_{\rm 1}$              &   $7.5{\pm}7$      &   --                &    --              &   $10{\pm}8$        &   $12{\pm}10$            \\
   FWHM                         &   $174{\pm}14$     &   --                &    --              &   $29{\pm}14$       &   $22{\pm}20$            \\
   Norm.                        &   $21.7{\pm}1.2$   &   --                &    --              &   $1.7{\pm}0.8$     &   $1.1{\pm}0.6$          \\
   rms ($\%$)                   &   $16.4{\pm}0.4$   &   --                &    --              &   $6.2{\pm}0.3$     &   $8.3{\pm}0.6$          \\
 \hline  \noalign{\smallskip}
   ${\nu}_{\rm 2}$              &   --               &   $48{\pm}13$       &    $36{\pm}11$     &   $82{\pm}19$       &   $77{\pm}25$            \\
   FWHM                         &   --               &   $147{\pm}15$      &    $166{\pm}12$    &   $137{\pm}25$      &   $160{\pm}24$           \\
   Norm.                        &   --               &   $17.2{\pm}1.5$    &    $9.1{\pm}0.5$   &   $1.3{\pm}0.4$     &   $1.3{\pm}0.4$          \\
   rms ($\%$)                   &   --               &   $16.4{\pm}0.7$    &    $16.5{\pm}0.5$  &   $6.5{\pm}0.3$     &   $8.4{\pm}0.6$          \\
 \hline  \noalign{\smallskip}
   ${\nu}_{\rm QPO}$            &   --               &   --                &    $231{\pm}5$     &   $226{\pm}8$       &   $233{\pm}4$            \\
   FWHM                         &   --               &   --                &    $16{\pm}11$     &   $33{\pm}18$       &   $17{\pm}16$            \\
   Norm.                        &   --               &   --                &    $0.46{\pm}0.15$ &   $0.16{\pm}0.06$   &   $0.12{\pm}0.04$        \\
   rms ($\%$)                   &   --               &   --                &    $3.8{\pm}2.1$   &   $2.3{\pm}1.0$     &   $2.7{\pm}1.8$          \\
 \hline  \noalign{\smallskip}
   ${\Gamma}_{\rm poisson}$     &   $0$              &    $0$              &       $0$          &   $0$            &     $0$                  \\
   ${\rm N}_{\rm poisson}$      &   $1.84{\pm}0.04$  &    $1.59{\pm}0.05$  &    $1.79{\pm}0.03$ &   $1.90{\pm}0.03$   &   $1.93{\pm}0.03$        \\
 \hline  \noalign{\smallskip}
   ${\chi}^{2}/{\nu}$           &   $0.84\,(213/252)$        &    $1.0\,(253/252)$        &    $0.98\,(243/249)$       &   $0.99\,(244/246)$      &   $0.80\,(196/246)$      \\
\noalign{\smallskip}
\hline  \noalign{\smallskip}
\multicolumn{6}{l}{$^{1}$ Model used: {\tt lorentz+lorentz+lorentz+powerlaw} for five {\it RXTE}/PCA observations.}\\
\multicolumn{6}{l}{$^{2}$ The PDS were created in the (2--60)\,keV energy band and $(8{\times}10^{-3}-4)$\,Hz frequency range.}\\
\multicolumn{6}{l}{$^{3}$ The frequency and width (${\nu}$ and FWHM) of the components are shown in units of mHz.}\\
\multicolumn{6}{l}{$^{4}$ Errors are $68\%$ confidence errors.}\\
\end{tabular}
\end{minipage}
\end{table*}

\subsubsection{Spectral analysis}

We fitted the brightest background-subtracted spectra with standard spectral models
in XSPEC 12.8.2 \citep{arnaud96}. All errors quoted in this work are $68\%$ ($1{\sigma}$) confidence. The spectral fits were limited to the 1-10, 4.5-28, 26-60\,keV range for the  
XRT, PCA and the HEXTE, respectively, where the calibration of the instruments is the best. The spectra were grouped 
in order to have at least 100, 50, 50\,counts for the XRT, PCA and HEXTE, respectively for each background-subtracted spectral channel and to avoid oversampling of the 
intrinsic energy resolution.

As it is a common procedure in observational studies of NS systems we used empirical models to describe the X-ray spectra. To compare our results with previous studies
we used the same (or very similar) spectral components. To fit the spectral continuum we used a model composed by photoelectric absorption (\citealt{balucinska92}; {\tt phabs} in XSPEC)
and a power-law with a high-energy exponential cutoff ({\tt cutoffpl}). The absorption component was not used in Obs.~1-4 and 11 (we did not fit the data from Obs.~12,13 because
of the lack of counts), since PCA 
spectra are not affected (i.e. at ${\rm E}{\ge}4.5$\,keV). In the case of XRT observations, with lower energy coverage (i.e. at ${\rm E}{\ge}1$\,keV) the absorption component 
was included in all our fits. This model provided significant wavy residuals in the whole spectra of all the observations with non acceptable chi-squared 
values (${\chi}^{2}$/${\nu}{\approx}5-60$; with ${\nu}$ being the number of degrees of freedom, i.e. d.o.f.). 

\begin{table*}
 \centering
 \begin{minipage}{120mm}
  \caption{Results from the ${\it RXTE}$ spectral analysis only$^{1}$.}
  \label{tab3}
  \begin{tabular}{@{}llllll@{}}
  \hline \noalign{\smallskip}
                                &   Obs.~1   &    Obs.~2   &    Obs.~3    &   Obs.~4    &   Obs.~11          \\
 \hline   \noalign{\smallskip}

   ${\rm E (1)}_{\rm gabs}$         & $7.6{\pm}2.8$        &  $8.0{\pm}2.3$     &  $9.0{\pm}0.5$     &  $6.5{\pm}0.7$      &  $5{\pm}2$    \\
   ${\sigma}{\rm (1)}$              & $4.6{\pm}1.6$        &  $4.1{\pm}0.8$     & $3.3{\pm}0.6$      &  $6{\pm}5$          &  $7{\pm}3$      \\
   ${\rm S (1)}$                    & $15{\pm}8$           & $9{\pm}7$          &  $2.2{\pm}0.9$     &  $70{\pm}40$        &  $11{\pm}10$      \\
 \hline   \noalign{\smallskip}
   ${\rm E (2)}_{\rm gabs}$         & $31.6{\pm}0.7$       & $31.3{\pm}0.8$     &  $31.8{\pm}0.8$    &  $29.7{\pm}1.2$     &  $28.4{\pm}1.0$   \\
   ${\sigma}{\rm (2)}$              & $4.3{\pm}0.4$        &  $4.6{\pm}0.4$     & $5.0{\pm}0.6$      &  $4.1{\pm}0.8$      &  $3.4{\pm}0.5$  \\
   ${\rm S (2)}$                    & $17{\pm}3$           & $20{\pm}4$         & $23{\pm}5$         &  $14{\pm}6$         &  $8{\pm}3$       \\
 \hline   \noalign{\smallskip}
   ${\rm E }_{\rm gauss}$           & $6.3{\pm}0.5$        & $6.40{\pm}0.10$    & $6.9{\pm}1.5$      &  $6.3{\pm}0.6$      &  $5.7{\pm}0.6$  \\
   ${\sigma}$                       & $0.18{\pm}0.20$      &   $0.22{\pm}0.20$  & $1.6{\pm}0.5$      &  $0.5{\pm}0.4$      &  $1.1{\pm}0.6$  \\
   ${\rm N}$                        & $0.012{\pm}0.005$    &  $0.009{\pm}0.004$ &  --                &  $0.014{\pm}0.010$  &  $0.02{\pm}0.01$ \\
 \hline   \noalign{\smallskip}
   ${\Gamma}_{\rm pow}$             & $-0.7{\pm}0.5$       & $-0.51{\pm}0.24$   & $-0.93{\pm}0.28$   &  $-0.7{\pm}0.5$     &  $-0.7{\pm}0.3$  \\
   ${\rm E}_{\rm cutoff}$           & $5.7{\pm}0.6$        & $6.6{\pm}0.5$      & $5.7{\pm}0.6$      &  $5.0{\pm}0.8$      &  $5.68{\pm}0.15$  \\
 \hline   \noalign{\smallskip}
   ${\rm F}_{\rm X}$                & $10{\pm}4$           & $8{\pm}4$          &  $4.7{\pm}0.1$     &  $4.5{\pm}0.3$      &  $1.3{\pm}0.7$      \\
   ${\rm L}_{\rm X}$                & $5.9{\pm}2.4$        & $4.7{\pm}2.4$      & $2.8{\pm}0.3$      & $2.7{\pm}0.3$       &  $0.8{\pm}0.4$  \\
 \hline   \noalign{\smallskip}
   ${\rm C}_{\rm PCA}$              & $1$                  & $1$                & $1$                &  $1$                &  $1$          \\
   ${\rm C}_{\rm HEXTE}$            & $0.50{\pm}0.05$      & $0.45{\pm}0.07$    & $1.3{\pm}0.3$      &  $1.5{\pm}0.3$      &  $0.40{\pm}0.07$   \\
 \hline   \noalign{\smallskip}
   ${\chi}^{2}$/${\nu}$             & $0.83\,(43/52)$      & $0.90\,(47/52)$    & $1.1\,(56/52)$     &  $1.2\,(64/52)$     &  $1.1\,(56/52)$  \\
\noalign{\smallskip}
\hline \noalign{\smallskip}
\multicolumn{6}{l}{\tiny{$^1$ Model used: {\tt constant${\times}$gabs${\times}$gabs( cutoffpl + gaussian )} for the 4.5-60\,keV {\it RXTE}} }\\
\multicolumn{6}{l}{\tiny{ spectra.} } \\
\multicolumn{6}{l}{\tiny{$^2$ Centroid, line-width of the lines, normalization of the {\tt gaussian} line, flux and luminosity} }\\
\multicolumn{6}{l}{\tiny{ in units of ${\rm keV}$, ${\rm cm}^{-2}\,{\rm s}^{-1}$, $10^{-9}\,{\rm erg}\,{\rm s}^{-1}\,{\rm cm}^{-2}$ and $10^{37}\,{\rm erg}\,{\rm s}^{-1}$, respectively. } }\\
\multicolumn{6}{l}{\tiny{$^3$ Errors are $68\%$ confidence errors.}}\\
\end{tabular}
\end{minipage}
\end{table*}

\begin{table*}
 \centering
 \begin{minipage}{180mm}
  \caption{Results from the ${\it Swift}$ $+$ {\it RXTE} spectral analysis$^{1}$.}
  \label{tab4}
  \begin{tabular}{@{}lcccccc@{}}
  \hline   \noalign{\smallskip}
                                &   Obs.~5   &    Obs.~6   &    Obs.~7    &   Obs.~8    &  Obs.~9   &   Obs.~10               \\
 \hline   \noalign{\smallskip}

   ${\rm N}_{\rm H}$\,$({\times}10^{22})\,({\rm cm}^{-2})$  &  $1.01{\pm}0.08$     &  $0.74{\pm}0.07$ &  $0.84{\pm}0.07$ &  $0.80{\pm}0.13$ & $0.77{\pm}0.07$  &  $0.93{\pm}0.12$           \\
   ${\rm E (1)}_{\rm gabs}$\,(${\rm keV}$)           &  $4.1{\pm}1.6$              &  $7.8{\pm}0.7$   &  $8.5{\pm}0.5$   & $7.0{\pm}1.0$    & $7.2{\pm}0.6$    &  $8.86{\pm}0.20$           \\
   ${\sigma}{\rm (1)}$\,(${\rm keV}$)                &  $6.6_{-0.5}^{+1.7}$        & $4.3{\pm}0.7$    & $3.0{\pm}0.8$    & $5.1{\pm}1.4$    & $5.5{\pm}0.6$    &  $7{\pm}3$                 \\
   ${\rm Strength (1)}$                              &  $18_{-5}^{+13}$            & $12{\pm}4$       & $1.9_{-0.4}^{+1.7}$  & $28{\pm}14$  & $15{\pm}5$       &  $60{\pm}30$               \\
 \hline  \noalign{\smallskip}
   ${\rm E (2)}_{\rm gabs}$\,(${\rm keV}$)           &  $30.8{\pm}0.8$             & $29.3{\pm}1.1$   & $26.8{\pm}0.9$   & $30.0{\pm}1.4$   & $30.0{\pm}2.0$   &  $27.5{\pm}2.3$            \\
   ${\sigma}{\rm (2)}$\,(${\rm keV}$)                &  $4.7{\pm}0.5$              &  $3.4{\pm}0.4$   & $2.7{\pm}0.5$    & $4.0{\pm}0.7$    & $3.5{\pm}0.5$    &  $2.15{\pm}1.0$            \\
   ${\rm Strength (2)}$                              &  $19{\pm}4$                 & $14{\pm}4$       & $6.0{\pm}2.0$    & $12{\pm}4$       & $16{\pm}9$       &  $10{\pm}8$                \\
 \hline  \noalign{\smallskip}
   ${\rm E }_{\rm gauss}$\,(${\rm keV}$)             &  $6.56{\pm}0.16$            & $6.23{\pm}0.20$  & $6.23{\pm}0.19$  & $6.15{\pm}0.4$   & $6.5{\pm}0.5$    &  $6.5{\pm}0.8$            \\
   ${\sigma}$\,(${\rm keV}$)                         &  $0.004_{-0.003}^{+0.23}$   & $0.4{\pm}0.3$    &  $0.34{\pm}0.20$ & $1.25{\pm}0.25$  & $0.003{\pm}0.003$    &  $1.4{\pm}0.9$             \\
   ${\rm N}$\,(${\rm cm}^{-2}\,{\rm s}^{-1}$)        &  $0.55{\pm}0.10$            & $0.007{\pm}0.004$  & $0.0013{\pm}0.0010$ &  $0.05{\pm}0.04$     & $0.0012{\pm}0.0007$ &  $0.07{\pm}0.05$      \\
 \hline  \noalign{\smallskip}
   ${\Gamma}_{\rm pow.}$                             &  $-0.02{\pm}0.21$           & $-0.27{\pm}0.12$   & $0.05{\pm}0.10$     & $-0.20{\pm}0.20$     & $-0.49{\pm}0.10$   &  $0.0{\pm}0.4$        \\
   ${\rm E}_{\rm cutoff}$\,(${\rm keV}$)             &  $5.4{\pm}0.4$              & $5.8{\pm}0.7$    &  $10.3{\pm}1.6$       & $6.0{\pm}0.5$        & $5.2{\pm}0.7$     &  $7{\pm}4$          \\
 \hline  \noalign{\smallskip}
   ${\rm F}_{\rm X}$\,$^1$\,(${\times}10^{-9}$\,${\rm erg}\,{\rm s}^{-1}\,{\rm cm}^{-2}$)   & $4.3{\pm}0.4$    &  $6.5{\pm}0.6$  &  $6.3{\pm}0.6$  &  $6.2{\pm}0.6$  &  $3.3{\pm}0.3$   &  $3.2{\pm}0.3$                          \\
   ${\rm L}_{\rm X}$\,(${\times}10^{37}$\,${\rm erg}\,{\rm s}^{-1}$) & $2.5{\pm}0.3$ & $3.8{\pm}0.4$ & $3.8{\pm}0.4$ & $3.7{\pm}0.4$ & $2.0{\pm}0.2$      &  $1.9{\pm}0.3$  \\
 \hline  \noalign{\smallskip}
   ${\rm C}_{\rm PCA}$                               &  $1$                        &  $1$                & $1$                &  $1$                & $1$                &  $1$            \\
   ${\rm C}_{\rm HEXTE}$                             &  $1.43{\pm}0.20$            &  $0.45{\pm}0.15$    & $0.11{\pm}0.05$    &  $0.33{\pm}0.06$    & $0.61{\pm}0.15$    &  $0.26{\pm}0.17$\\
   ${\rm C}_{\rm XRT}$                               &  $0.87{\pm}0.08$            &  $0.39{\pm}0.03$    & $0.39{\pm}0.03$    &  $0.29{\pm}0.03$    & $0.41{\pm}0.03$    &  $0.48{\pm}0.05$\\
 \hline  \noalign{\smallskip}
   ${\chi}^{2}$/${\nu}$                              &  $0.93\,(182/195)$          &  $0.97\,(152/156)$  & $1.13\,(183/161)$  &  $0.87\,(97/112)$   & $0.94\,(97/103)$   &  $1.02\,(99/97)$\\
\noalign{\smallskip}
\hline \noalign{\smallskip}
\multicolumn{7}{l}{\tiny{$^1$ Model used: {\tt phabs${\times}$constant${\times}$gabs${\times}$gabs( cutoffpl + gaussian )} for the 1-60\,keV {\it Swift} $+$ {\it RXTE} spectra.}} \\
\multicolumn{7}{l}{\tiny{$^2$ Errors are $68\%$ confidence errors.}} \\
\end{tabular}
\end{minipage}
\end{table*}

The fits and the residuals of the whole spectra improved (${\chi}^{2}$/${\nu}{\approx}1.5-6$) by the addition of a cyclotron line scattering feature (CRSF; hereafter referred 
to as ``cyclotron feature''). It was accounted for with a gaussian absorption line ({\tt gabs}). The use of this model component and/or the obtained line parameters are consistent with
previous studies \citep{kreykenbohm05,tsygankov06,reig06,reig13,lutovinov15}. The parameters were let to be 
free in the HEXTE data and tied to those in the XRT and PCA data. 

The fits still show significant residuals at 7-10\,keV. This feature is known as the ``10 keV feature'', and its origin is uncertain \citep{coburn02}. We modeled it as a 
gaussian absorption line ({\tt gabs}), thus improving the fits substantially (${\chi}^{2}$/${\nu}{\approx}1.0-1.2$). In this case the parameters were let to be free in the PCA data and tied to 
those in the XRT and HEXTE data. The values obtained are within those expected \citep{coburn02} but with broader dispersion than those obtained e.g. for XTE~J1946+274 by \citet{muller12}.

Although the beneath of the spectral fits, there are still peaked positive residuals in the region 6-7\,keV of the PCA spectra. We accounted for them adding a gaussian line profile
({\tt gaussian}) at 6.4-6.97\,keV in order to account for the Fe K$_{\alpha}$ fluorescence line. The PCA residuals flattened and we obtained the final fitting solution in this way.

\begin{figure}
\centering
 \includegraphics[bb=0 30 612 370,width=8.7cm,angle=270,clip]{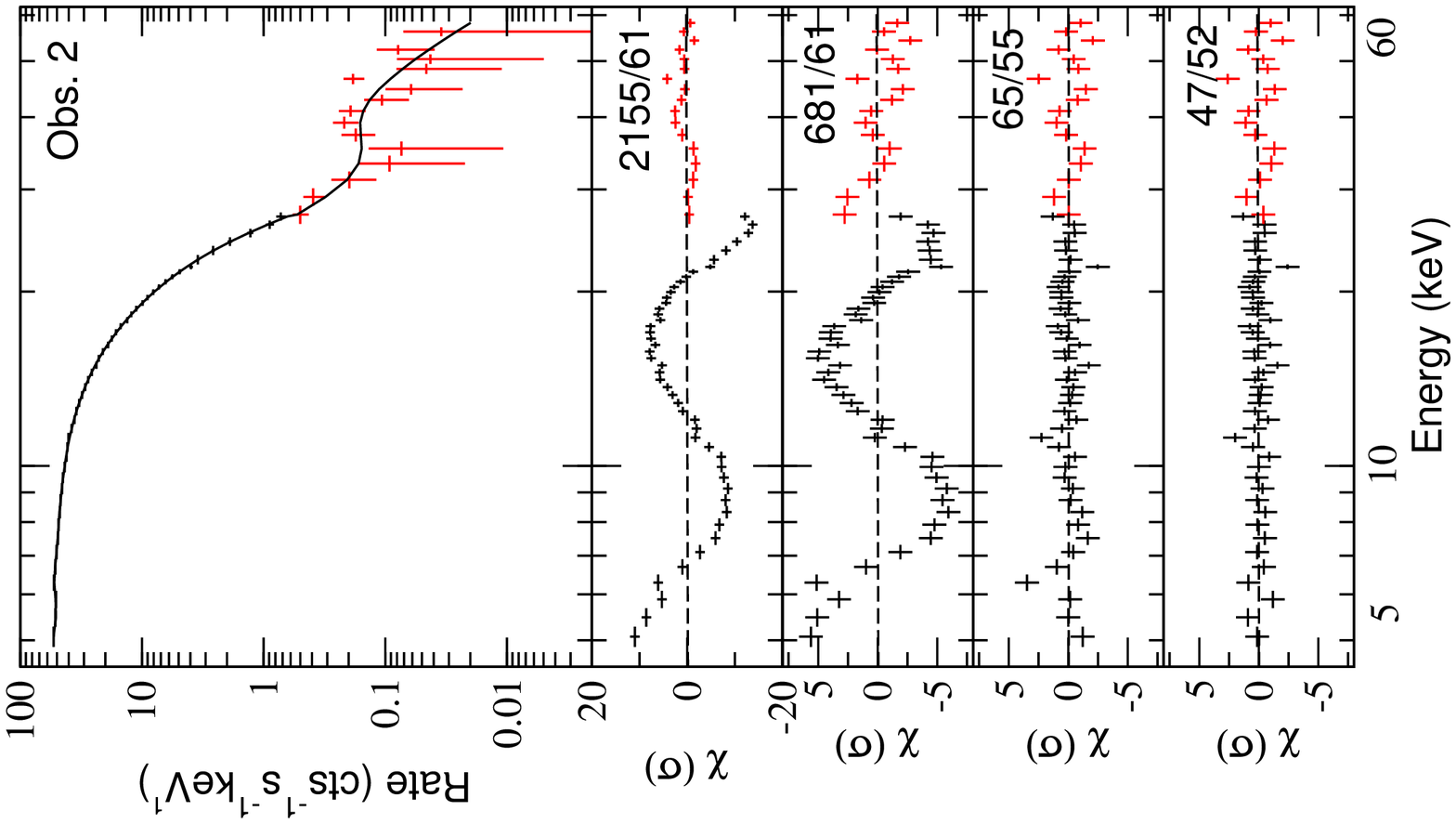}
 \includegraphics[bb=0 90 612 372,width=8.7cm,angle=270,clip]{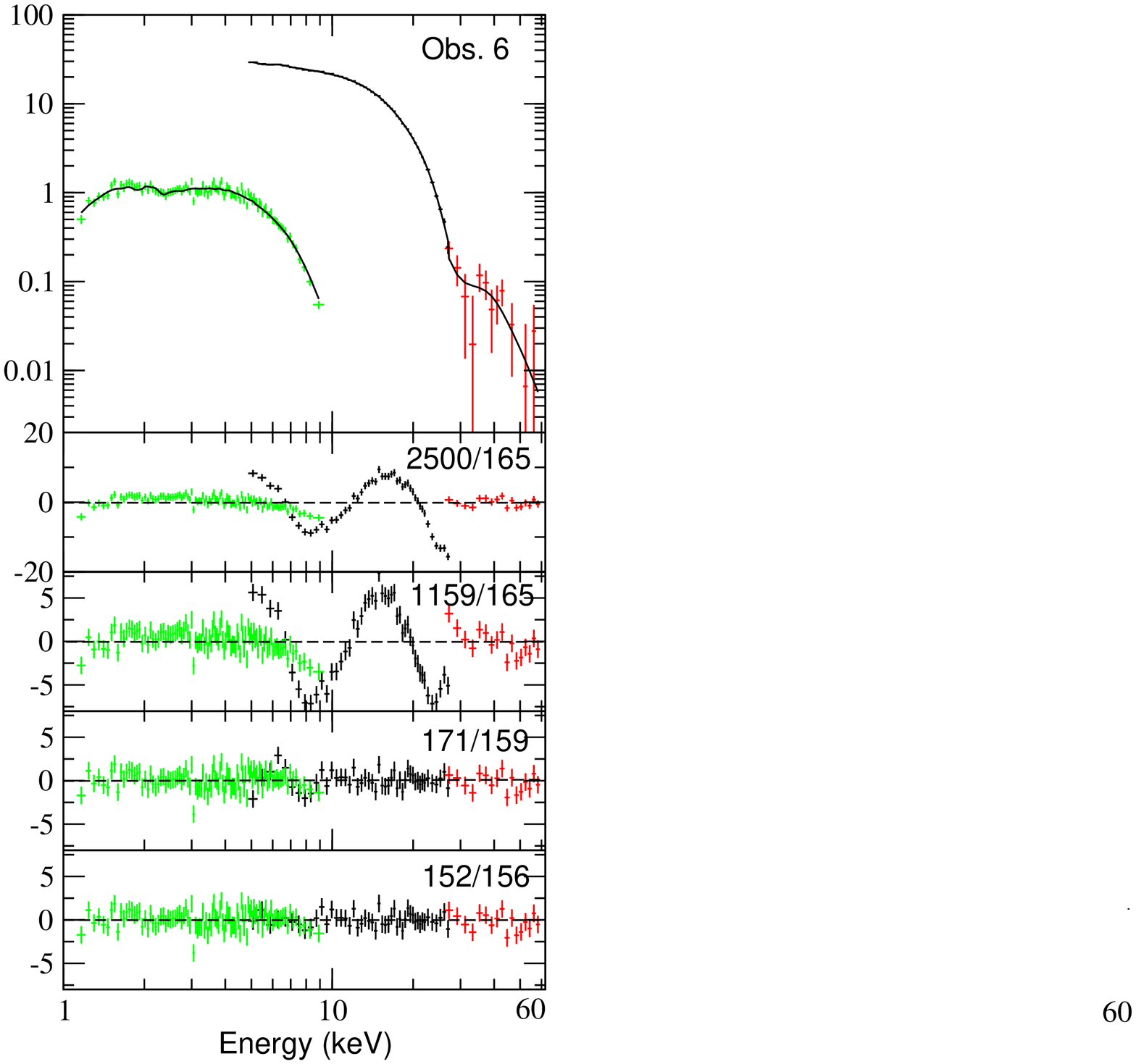}
 \caption{The two (upper) panels (left and right) show two example spectra of {\vzero}. Obs.~2 is one of the two observations with the highest count rates while Obs.~6 is the dataset
with the highest count rate with all the instruments available (black, red and green correspond to PCA, HEXTE and XRT, respectively). The (lower) panels show the behaviour of the
residuals when adding the different components one by one. Number at the right of each panel indicate the best fit statistics (${\chi}^{2}$/d.o.f.). From upper to lower panels: 1)
best-fit using the {\tt cutoffpl} model component only (i.e. continuum model); 2) residuals after adding the ${\approx}10$\,keV (absorption) feature to the continuum model; 3) residuals after
adding the cyclotron feature; 4) residuals after adding the Fe ${\rm K}_{\alpha}$ line to the model (and final residuals of the model). Note that the same vertical axis is used
in both panels.
}
 \label{fig4}
\end{figure}

To account for uncertainties in the absolute flux normalization between PCA and HEXTE we introduced a multiplicative constant that was fixed to 1 for the PCA and let to vary freely for
HEXTE. In the case of XRT, PCA and HEXTE data we fixed the multiplicative constant of PCA to 1 and the rest to vary freely. The most relevant results of this spectral analysis and 
the derived unabsorbed fluxes and luminosities are shown in Tab.~\ref{tab3},\ref{tab4} and in Fig.~\ref{fig4}.

\begin{figure}
\centering
 \includegraphics[bb=39 22 564 700,width=6.5cm,angle=270,clip]{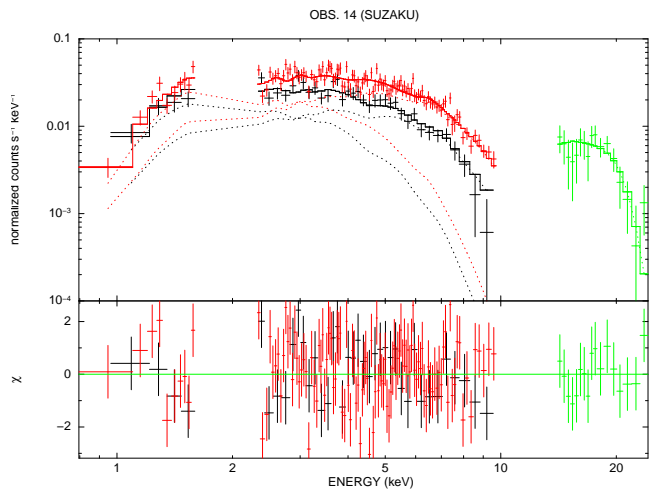}
 \caption{Energy spectrum from V~0332+53 during Obs.~14 with {\it Suzaku} (see Tab.~\ref{tab5} for the observation details and the results of the
spectral analysis). Black, red and green correspond to XIS1, XIS3 and PIN, respectively. }
 \label{fig5}
\end{figure}

In addition, we fitted the spectra obtained with {\it Suzaku}, during much a lower luminosity period (see Fig.~\ref{fig5} and Tab.~\ref{tab5}). Spectra from XIS1, XIS3 and PIN were fitted in
the 0.8-10,14-25\,keV energy range, with the 1.6-2.3\,keV energy range ignored due to the  uncertainties in calibration associated with the instrumental Si K edge\footnote{ http://heasarc.gsfc.nasa.gov/docs/suzaku/analysis/\\sical.html}. The
spectra were grouped to have at least 50,250\,counts for the two XIS and PIN, respectively. In this case a model composed by photoelectric absorption ({\tt phabs}) and
a power-law with an exponential high-energy cutoff of the type ``{\tt highecut}'' was a rather better description than ``{\tt cutoffpl}'' (i.e. ${\Delta}\,{\chi}^{2}=15$ for the
same number of d.o.f and better residuals). With this model we obtained a statistics of ${\chi}^{2}$/${\nu}=1.4$ and systematic residuals at ${\rm E}<4$\,KeV, that we improved by adding
a blackbody spectrum with normalization proportional to the surface area ({\tt bbodyrad}), obtaining a fit statistics of (${\chi}^{2}$/${\nu}=1.3$ for the same number of d.o.f.). Because
the normalization of the black-body component was not well constrained we tried to improve it and changed it by an emission component describing an accretion disc consisting of
multiple blackbody components ({\tt diskbb}; \citealt{mitsuda84,makishima86}). We obtained a very similar fit statistics (see Tab.~\ref{tab5}). A plausible origin for
the first model component ({\tt bbodyrad}) is the NS surface or perhaps the boundary layer (if present). Low-level accretion onto the surface of a neutron star can indeed
produce a black-body like spectrum \citep{zampieri95}. Still, a disc origin cannot be ruled out since it was also possible to fit the data with
a multicolor disc blackbody. Similar results for a set of low-accreting NS have been recently obtained \citep{armaspadilla13}, but the temperatures found 
are ${\rm kT}{\approx}=0.5-0.7$\,keV, thus much lower than in our case (${\rm kT}{\approx}=2.4{\pm}0.9$\,keV). The higher value we obtain might be indicative of 
the much higher accretion rate in our case. Nevertheless, with the current data it is not possible to discern which model component is better to describe the data.

To account for the uncertainties in the absolute flux normalization between XIS1, XIS3 and PIN we introduced a multiplicative constant that was fixed to 1 for XIS1 and let to vary
freely for XIS3 and PIN. The most relevant results of this spectral analysis and the derived unabsorbed fluxes and luminosities are shown in Tab.~\ref{tab5}.

\begin{table}
 \centering
 \begin{minipage}{100mm}
  \caption{Results from the ${\it Suzaku}$ spectral analysis$^{1}$.}
  \label{tab5}
  \begin{tabular}{@{}lc@{}}
  \hline  \noalign{\smallskip}
                                &   Obs.~14   \\
 \hline  \noalign{\smallskip}
   ${\rm N}_{\rm H}$\,$({\times}10^{22})$\,(${\rm cm}^{-2}$)       &   $1.2{\pm}0.1$    \\
   ${\rm kT}_{\rm bb}$\,(${\rm keV}$)                &   $2.4{\pm}0.9$    \\
   ${\rm N}_{\rm bb}$                                &   $0.011{\pm}0.007$    \\
 \hline  \noalign{\smallskip}
   ${\Gamma}_{\rm pow.}$                             &   $-1.3{\pm}1.6$    \\
   ${\rm E}_{\rm cutoff}$\,(${\rm keV}$)             &   $15-190$    \\
   ${\rm E}_{\rm f}$\,(${\rm keV}$)                  &   $1.0{\pm}1.0$    \\
 \hline  \noalign{\smallskip}
   ${\rm F}_{\rm X}$\,(${\times}10^{-11}$)\,(${\rm erg}\,{\rm s}^{-1}\,{\rm cm}^{-2}$)           &        $2.4{\pm}0.9$ \\
   ${\rm L}_{\rm X}$\,(${\times}10^{35}$)\,$({\rm erg}\,{\rm s}^{-1})$            &        $1.4{\pm}0.5$ \\
 \hline  \noalign{\smallskip}
   ${\rm C}_{\rm XIS1}$                              &   $1$                   \\
   ${\rm C}_{\rm XIS3}$                              &   $1.35{\pm}0.04$     \\
   ${\rm C}_{\rm PIN}$                               &   $0.4{\pm}0.3$     \\
 \hline  \noalign{\smallskip}
   ${\chi}^{2}$/${\nu}$                              &   $1.35\,(221/163)$     \\
\noalign{\smallskip}
\hline  \noalign{\smallskip}
\multicolumn{2}{l}{\tiny{$^1$ Model used: {\tt phabs${\times}$constant${\times}$highecut(diskbb + }}  }\\
\multicolumn{2}{l}{\tiny{{\tt powerlaw)} for the 0.8-30\,keV {\it Suzaku} spectrum. } }\\
\multicolumn{2}{l}{\tiny{$^2$ Errors are $68\%$ confidence errors. } }\\
\end{tabular}
\end{minipage}
\end{table}

\section{Discussion}  \label{discuss}

\subsection{Long-term Be-disc/neutron star interaction }

From the long-term observations of {\vzero} we can see that {\vzero} spends most of its time in the optical brightening phases (Fig~\ref{fig7}). The most 
significant one is the period between 1993-2005. It started approx. 4 years after the giant outburst that the system exhibited in 1989 and peaked simultaneously with the X-ray maximum
in 2004. Indeed all of the optical brightening episodes of BQ~Cam are accompanied with the X-ray activities. It is important to note that the brightness of the optical companion
decreases after the X-ray outbursts. This behavior can be well explained by the weakening of the decretion disc during the mass transfer to the NS.

Having a moderate eccentricity of ${\rm e}=0.31$ with a short orbital period of ${\rm P}_{\rm orb}=34.25$\,d, {\vzero} is one of the BeXBs for which the effect of the 
truncation exerted by the NS is likely to be observed. Due to the small orbit of the companion, the decretion disc cannot expand so much. Thus, it is truncated at a radius 
smaller than the critical Roche radius. Since the amount of the material supported by the Be star to the disc would be accumulated in time, the increase in the optical 
brightness can be understood in this way. However the accumulation of the material particularly at the outer part of the disc makes these regions to become optically 
thick. According to theory by either the radiation-driving warping or a global density wave the outer part of the disc is strongly elongated 
\citep{okazaki01}. In general, the existence of the global density waves in the decretion disc are observationally supported by the V$/$R variations seen in the 
emission line profiles. Although the line profile variations of {\vzero} were previously observed \citep{negueruela98} during
the period 1990-1991, ${\rm H}_{\alpha}$ lines are always seen in nearly symmetric single-peaked emissions without any significant variations for the
time interval 2006-2014. Therefore, we do not have any evidence to support the idea of the perturbations occurred in the disc as well as the V/R variations.

\subsection{X-ray behaviour}  

\subsubsection{Spectral evolution}  

From the {\it RXTE} and {\it Swift} observations reported in this work we see a decrease of the X-ray luminosity of {\vzero} during the latest part of the 
intermediate-luminosity 2008 outburst. Assuming a 7\,kpc distance \citep{negueruela99}, the luminosity gradually dropped
from $\sim$6 to $\sim$1$\times10^{37}$\,erg\,s$^{-1}$ in the 5-60\,keV energy range from Obs.~1-11. The luminosities are also
in accordance with those obtained during the decay phase of the 2005 outburst \citep{mowlavi06}. Eventually, during the {\it Suzaku} observation (Obs.~14) we obtain the
lowest luminosity ($\sim$1$\times10^{35}$\,erg\,s$^{-1}$). 

A recent study of {\vzero} during giant outbursts at different luminosities has provided crucial insights into the most frequent
states of {\vzero} \citep{reig13}. Their Fig.~2 shows
the existence of two separate regimes (the high-luminosity diagonal branch and the low-luminosity horizontal branch). They are
separated by a critical luminosity (${\rm L}_{\rm crit}{\approx}1-4{\times}10^{37}$\,erg\,s$^{-1}$), which corresponds to the range of luminosities 
reported in our work during Obs.~1-11. Both regimes/states
are clearly well-populated by X-ray observations whilst observations around the critical luminosity are scarce.  

The spectral variability of {\vzero} has been well studied by different authors, confirming the softening of {\vzero} as the flux decreased. \citet{baykal07} suggested 
that this type of spectral softening with decreasing flux was mainly a consequence of mass accretion rate change. In particular, \citet{reig10} (and references therein) explained 
that increased mass accretion rates are expected to result in harder X-ray spectra due to Comptonization processes. In our work we see the same trend photon 
index versus flux. 

\begin{figure}
\centering
 \includegraphics[bb=0 0 612 792,width=7.2cm,angle=270,clip]{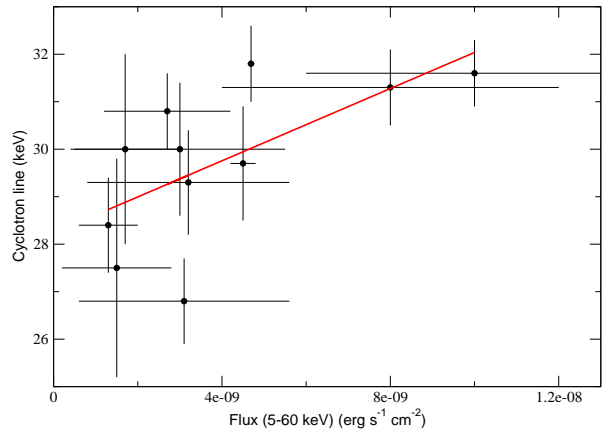}
 \caption{Flux in the 5-60\,keV energy range versus the cyclotron line energy centre for Obs.~1-11. See on-line material Tab.~\ref{otab3} for the (not-rounded) list of
values used. }
 \label{fig8}
\end{figure}

The values previously obtained for the energy centre of the cyclotron line have shown a strong anti-correlation with luminosity in {\vzero}
\citep{tsygankov06,mowlavi06,tsygankov10,reig13} during giant outbursts. Somehow weaker anti-correlation has been found in 4U~0115+63 \citep{mihara04,nakajima06} 
and no correlation in 1A~0535+262 \citep{caballero07}. In our work a positive correlation of the line energy centre with luminosity is observed (see Fig.~\ref{fig8}). The correlation 
coefficient of the fit is +0.64 (i.e. null two-tailed probability of 0.03). This kind of behaviour has also been observed in A~0535+262 during an intermediate-luminosity 
outburst (${\rm L}_{\rm X}=4{\times}10^{37}\,{\rm erg}\,{\rm s}^{-1}$; \citealt{sartore15}). We notice that previously \citet{reig13} already pointed out for a tentative positive
correlation for the latter source. This effect appears to be a natural consequence of accretion onto 
X-ray pulsars with the strongest magnetic fields in the sub-critical regime \citep{mushtukov15}. The values we have obtained for the width of the cyclotron line are 
all ${\approx}4$\,keV, in contrast to the values obtained during high flux states 
(6-8\,keV; \citealt{tsygankov10}) but in full agreement with those obtained in intermediate flux states (4\,keV; \citealt{mowlavi06}).

\subsubsection{Quasi Periodic Oscillations}  

We have found that the PDS of {\vzero} during our observations are best described by band-limited noise plus QPO noise during the latest observations. The QPO noise consists of 
a component at a frequency of ${\approx}0.22-0.23$\,Hz. This feature has also been reported in the 2-60\,keV {\it RXTE}-PCA PDSs obtained during 
the 2004-2005 outburst decay \citep{qu05}. They found a QPO at ${\approx}$0.22\,Hz in addition to  
a narrow peak (i.e. pulsation from the pulsar) close to that frequency (${\approx}$0.23\,Hz). The QPO was detected when the luminosity was at a mid-flux or even lower flux level of
the outburst. The rms strength of this QPO also amounts to a few per cent in that work. The 0.22\,Hz QPO was observed around 
the spin frequency in both the {\it INTEGRAL} \citep{mowlavi06} and {\it RXTE} data during the 2005 outburst decline \citep{qu05,reig06}. In addition, \citet{takeshima94} 
discovered another QPO with a centroid frequency of ${\approx}$0.05\,Hz and a relative root mean square (rms) amplitude of ${\approx}$5\%, in the 2.3-37.2\,keV data from the 
1989 outburst of {\vzero}. We do not find any signal of the QPO at ${\approx}$0.05\,Hz although
we find the presence of a broad component at around that frequency (that may turn eventually into peak noise, i.e. QPO). The QPO detected in 
{\vzero} at ${\approx}$0.22\,Hz constitutes the first detection of a QPO riding on the spin frequency of a NS and points out to a strong coupling between the periodic
(i.e. the pulses) and the red-noise (i.e. broad band QPO signal) components \citep{qu05}.

The existence of a QPO signal with a similar frequency to the spin of the NS suggests that they are related somehow. One model to explain
such coincidence was given by \citet{lazzati97} and \citet{burderi97}. They proposed that the QPO is originated in the magnetosphere around the NS. Additionally, they 
argued that the accretion flow is inhomogeneous near the surface of the NS, and that
random shots are characterized by an arbitrary degree of modulation. The combination of rotation and radiative transfer effects should therefore produce a periodic modulation 
of the shots similar to that of any continuum X-ray emission from the polar caps. They proposed that a coupling between the periodic and red-noise variability should be frequently 
present in X-ray pulsars. 

In our case and in contrast to previous observations during giant flares \citep{reig06,reig08} the QPO signal appears during the lowest of the measured fluxes
(${\rm L}_{\rm X}{\le}4{\times}10^{37}$\,${\rm erg}\,{\rm s}^{-1}$). If the QPO has an origin on the magnetosphere of 
the NS (and is not an artifact from our observations), it should be ``more visible'' when the accretion column is more tenuous (or even absent). Summarizing, it looks 
like that during our latest Obs.~6-11 we might be witnessing the innermost parts closest to the magnetosphere surroundings.

\section{Summary and conclusions}  \label{conclus}

In this paper we present a multi-wavelength study of the Be/X-ray binary system {\vzero} with the main goal of studying its transient behavior during its intermediate 
luminosity X-ray outburst in 2008. After 
being ${\approx}3.5$\,yr in quiescence in X-rays the system underwent a new active period. A 
new X-ray outburst was detected around 17 October 2008, roughly six days before the periastron passage of the NS. We note that the optical companion entered the brightening 
state nearly 1.5\,yr before the X-ray outburst. 

{\vzero} was in quiescence until the optical magnitude approached a peak value of ${\rm V}=15.2$\,mag. at the end of 2009, indicating 
the beginning of another mass ejection event. In November 2009 a series of five small Type I periodic outbursts was observed. 
The X-ray activities finished by the end of May 2010 whist the optical magnitude was still fading. It 
ceased around August 2011 for about ${\approx}3.5$\,yr. Our recent data showed that a mass ejection event is currently taking place, exhibiting an increase 
of ${\approx}0.6$ in the V mag. It probably started around 2013 and 
peaked in June 2015 when a new (type II) X-ray outburst was detected. This is similar to those observed during the brightening episodes occurred e.g. in 1983 and 
in 2004, but higher than the ones in 2008 and 2009. Our only IR measurement from December 2014 also showed an unusually bright Be companion. 


The broad-band 1-60\,keV X-ray spectra of the NS during the decay of the 2008 outburst were all well fitted with the standard {\tt cutoffpl} phenomenological 
model, enhanced by a narrow iron ${\rm K}_{\alpha}$ fluorescence line at 6.4\,keV together with a cyclotron line scattering feature at ${\approx}30$\,keV. Our study 
confirms the softening of {\vzero} as the flux decreased. During our lowest flux observation made with 
Suzaku (${\rm L}_{\rm X}=10^{35}\,{\rm erg}\,{\rm s}^{-1}$) on 2010, compatible 
with {\vzero} being at a level slightly above from quiescence, we detected a very soft spectrum that could be well described by adding a (single or multiple) soft black-body 
component(s). 

We tentatively see an increase of the cyclotron line energy with increasing flux. If confirmed (with better observations) this might constitute the first detection
of such positive correlation of the cyclotron line energy versus luminosity in {\vzero}. Regarding the fast aperiodic timing properties, we 
detect a QPO at $227{\pm}9$\,mHz during the lowest luminosities. The latter might indicate that the inner regions surrounding the 
magnetosphere are more visible during the lowest flux stages. Due to their low significance we advice that longer and better observations 
with current (and/or future) X-ray satellites are needed in order to confirm/discard these results.

\vspace{5mm}

\textbf{Acknowledgments}. We thank the anonymous referee and S. Campana for helpful comments. We also thank K. Page and the {\it Swift} team at Leicester (UK), A.~J. 
Castro-Tirado (on behalf of the BOOTES collaboration), I. E. Papadakis and J. Garcia-Rojas for helpful discussions/insights and/or for making part of these observations 
possible. RH acknowledges GA CR grant 13-33324S. MCG acknowledges support by the European social fund within the 
framework of realizing the project ``Support of inter-sectoral mobility and quality enhancement of research teams at Czech Technical University in Prague",
CZ.1.07/2.3.00/30.0034. This research has made use of the General High-energy Aperiodic Timing Software (GHATS) package developed by T.M. Belloni at INAF - Osservatorio Astronomico 
di Brera.


\clearpage

\pagebreak

\setcounter{page}{0}
\setcounter{table}{0}
\setcounter{figure}{0}

\begin{table}
  \caption{\textbf{ON-LINE MATERIAL}. Optical photometric observations from the
IAC80 telescope.}
  \label{otab1}
\begin{center}
\begin{tabular}{lcccc}
\hline\hline
MJD       &        V    &   error  &   B   &   error   \\
\hline \hline
56871.2368550198 &  15.460 & 0.013  &  17.037   &  0.071    \\
56993.9640124198 &  15.360 & 0.006  &  16.919   &  0.021    \\
57009.0723979599 &  15.373 & 0.005  &  17.006   &  0.013    \\
57026.9773657001 &  15.372 & 0.008  &  16.945   &  0.031    \\
57036.8705536202 &  15.373 & 0.005  &  16.967   &  0.013    \\
57054.0418117600 &  15.332 & 0.008  &  16.853   &  0.031    \\
57054.8470330602 &  15.364 & 0.006  &  16.955   &  0.024    \\

\hline
\end{tabular}
\end{center}
\end{table}

\begin{table*}
  \caption{\textbf{ON-LINE MATERIAL}. ${\rm H}_{\alpha}$ line EW and FWHM measurements of {\vzero}. Negative values of EW indicate that the line is in emission.}
  \label{otab2}
\begin{center}
\begin{tabular}{lcccc} 
\hline\hline
DATE      &  MJD       & EW & FWHM  &Telescope\\
&&(\AA)  &  (\AA) & \\
\hline \hline
26-Sep-2006  &54004.9238 &$-$6.21$\pm$0.71 &6.18$\pm$0.38 &RTT150 \\
23-Dec-2006  &54092.9543 &$-$7.05$\pm$0.56 &5.83$\pm$0.55 &RTT150\\
07-Apr-2007  &54197.7388 &$-$5.85$\pm$0.33 &6.48$\pm$0.83 &RTT150\\
19-Jul-2007  &54300.0279 &$-$5.11$\pm$0.10 &6.83$\pm$0.22 &RTT150\\
20-Jul-2007  &54301.0591 &$-$4.80$\pm$0.49 &6.40$\pm$0.68 &RTT150\\
13-Sep-2007  &54356.9962 &$-$3.82$\pm$0.18 &6.20$\pm$0.67 &RTT150\\
06-Oct-2007  &54379.0510 &$-$5.26$\pm$0.18 &6.62$\pm$0.40 &RTT150\\
14-Dec-2007  &54448.8768 &$-$4.62$\pm$0.17 &6.42$\pm$0.29 &RTT150\\
26-Aug-2008  &54704.9757 &$-$4.70$\pm$0.09 &6.31$\pm$0.48 &RTT150\\
22-Sep-2008  &54731.0652 &$-$3.97$\pm$0.09 &6.49$\pm$0.62 &RTT150\\
29-Oct-2008  &54768.8500 &$-$4.45$\pm$0.10 &6.69$\pm$0.26 &RTT150\\
04-Mar-2009  &54894.7260 &$-$3.13$\pm$0.23 &6.36$\pm$0.34 &RTT150\\
24-Sep-2009  &55098.9814 &$-$5.21$\pm$0.12 &6.56$\pm$0.44 &RTT150\\
21-Oct-2011  &55855.9805 &$-$6.44$\pm$0.10 &7.84$\pm$0.19 &OSN-150\\
16-Jan-2012  &55942.8952 &$-$4.38$\pm$0.07 &7.22$\pm$0.28 &RTT150\\
17-Jan-2012  &55943.7739 &$-$4.86$\pm$0.07 &9.38$\pm$0.62 &RTT150\\
31-Jan-2012  &55957.8646 &$-$4.08$\pm$0.11 &7.97$\pm$0.48 &OSN-150\\
31-Jan-2012  &55957.8864 &$-$4.78$\pm$0.22 &9.15$\pm$0.49 &OSN-150\\
05-Feb-2012  &55962.8764 &$-$5.45$\pm$0.10 &7.42$\pm$0.20 &RTT150\\
04-Mar-2012  &55990.8104 &$-$4.22$\pm$0.09 &6.82$\pm$0.17 &RTT150\\
27-Mar-2012  &56013.8502 &$-$3.61$\pm$0.23 &7.78$\pm$0.45 &OSN-150\\
27-Mar-2012  &56013.8618 &$-$4.31$\pm$0.31 &8.49$\pm$0.49 &OSN-150\\
29-Jul-2012  &56137.2382 &$-$4.84$\pm$0.30 &7.33$\pm$0.57 &NOT\\
08-Sep-2012  &56178.1507 &$-$6.56$\pm$0.32 &10.05$\pm$0.06 &OSN-150\\
09-Sep-2012  &56179.2028 &$-$4.79$\pm$0.20 &6.77$\pm$0.57 &NOT\\
16-Sep-2012  &56186.0446 &$-$5.83$\pm$0.10 &6.84$\pm$0.28 &RTT150\\
10-Oct-2012  &56210.1002 &$-$6.14$\pm$0.18 &7.01$\pm$0.03 &OSN-150 \\
09-Nov-2013  &56605.0347 &$-$7.44$\pm$0.08 &13.73$\pm$0.66 &RTT150\\
05-Jan-2014  &56662.8496 &$-$5.25$\pm$0.14 &6.48$\pm$0.67 &RTT150\\
05-Feb-2014  &56693.7630 &$-$4.34$\pm$0.20 &6.62$\pm$0.37 &RTT150\\
06-Feb-2014  &56694.8018 &$-$4.93$\pm$0.33 &7.28$\pm$0.40 &RTT150\\
20-Oct-2014  &56951.0575 &$-$4.92$\pm$0.07 &6.96$\pm$0.48 &RTT150\\
21-Oct-2014  &56951.0563 &$-$4.81$\pm$0.11 &6.32$\pm$0.13 &RTT150\\
22-Nov-2014  &56983.9104 &$-$5.23$\pm$0.13 &7.27$\pm$0.20 &RTT150\\
\hline
\end{tabular}
\end{center}
\end{table*}

\clearpage

\pagebreak

\begin{table}
 \centering
 \begin{minipage}{90mm}
  \caption{\textbf{ON-LINE MATERIAL}. Not-rounded values from Tab.~\ref{tab3} used in Fig.~\ref{fig8}}
  \label{otab3}
  \begin{tabular}{@{}lll@{}}
  \hline \noalign{\smallskip}
       Obs.                     &   ${\rm E (2)}_{\rm gabs}$\,(${\rm keV}$)   &   ${\rm F}_{\rm X}$\,(${\rm erg}\,{\rm s}^{-1}\,{\rm cm}^{-2}$)            \\
\noalign{\smallskip}
 \hline   \noalign{\smallskip}
   1    &   31.6${\pm}$0.7   &   (1.0${\pm}$0.4)${\times}10^{-8}$     \\
   2    &   31.3${\pm}$0.8   &   (0.8${\pm}$0.4)${\times}10^{-8}$    \\
   3    &   31.8${\pm}$0.8   &   (4.70${\pm}$0.10)${\times}10^{-9}$   \\
   4    &   29.7${\pm}$1.2   &   (4.5${\pm}$0.3)${\times}10^{-9}$    \\
   5    &   30.8${\pm}$0.8   &   (2.7${\pm}$1.5)${\times}10^{-9}$    \\
   6    &   29.3${\pm}$1.1   &   (3.2${\pm}$2.4)${\times}10^{-9}$    \\
   7    &   26.8${\pm}$0.9   &   (3.1${\pm}$2.5)${\times}10^{-9}$    \\
   8    &   30.0${\pm}$1.4   &   (3.0${\pm}$2.5)${\times}10^{-9}$    \\
   9    &   30.0${\pm}$2.0   &   (1.7${\pm}$1.3)${\times}10^{-9}$    \\
  10    &   27.5${\pm}$2.3   &   (1.5${\pm}$1.3)${\times}10^{-9}$    \\
  11    &   28.4${\pm}$1.0   &   (1.3${\pm}$0.7)${\times}10^{-9}$    \\
\noalign{\smallskip}
\hline \noalign{\smallskip}
\multicolumn{3}{l}{\tiny{$^1$ Fluxes are in the 5-60\,keV energy range.}}\\
\multicolumn{3}{l}{\tiny{$^2$ Errors are $68\%$ confidence errors.}}\\
\end{tabular}
\end{minipage}
\end{table}

\clearpage

\pagebreak

\end{document}